# Unpacking Cultural Perceptions of Future Elder Care through Design Fiction

Ng, Tse Pei*[a]; Lee, Jung-Joo [a]; Wu, Yiying[b]

[a] National University of Singapore, Singapore, Singapore
[b] The Hong Kong Polytechnic University, Hung Hom, Hong Kong
* ngtsepei@gmail.com

We present a case using Design Fiction to unpack cultural perceptions of future elder care rooted in the Asian context of Singapore. We created two design fictions, addressing the tensions between filial piety and automated care and the controversy of integrating elder care facilities into residential communities. The design fictions took the visual forms of a shopping web page and a petition site and the public were invited to make fictional decisions. Received in total 109 responses, we identify the key tensions and value conflicts and illustrate them through visual narratives. Further, we propose the Asian perspective of positioning relationships as the protagonist in creating elder care design fiction.

## 1 Introduction

Singapore faces a rapidly aging population, foreseeing that half its population will be 65 years old and above by 2050 (United Nations, Department of Economic and Social Affairs, Population Division 2017). While traditional nursing homes and elder care models have followed clinical typologies (Wong, Pang & Yap 2014), the Singapore government is exploring future models of elder care, focusing on enhancing care efficiency through automated technologies and integrating elder care facilities in residential communities.

The background of this paper is a government-funded project, Designing Future-ready and Sustainable Nursing Homes for Person-Centric Care Models in Communities, conducted from 2017 to 2020. The project explores future typologies of nursing homes in Singapore, which aim to be people-centred, sustainable and future ready. Around the project, a few conflicts were observed around the innovation of the care process through automated technologies and community integration of the elder care. They are critical tensions between the Confucian value of filial piety and automated care, and between the residence community and the elder care facilities.

To unpack these underlying tensions and value conflicts deeply rooted in Singapore's culture, we created two pieces of Design Fiction as a method to engage the public in the discussion, helping them articulate opinions and debate what kinds of future they desire in elder care. We developed design fictions by tweaking the original design proposals for the government agencies in the project. The design fictions were then published to the local public to provoke their reflections and responses. In this paper, we introduce the development of Design Fictions in the topic of the future elder care in Singapore, and thematic analysis of the community responses. The themes reveal cultural perceptions and value conflicts around the future elder care in Singapore.

## 2 Design fiction
### 2.1 Engaging in Design Fiction

Design fiction is gaining popularity as a design and research tool for technological innovation in the Human-Computer Interaction community. Researchers often refer a design fiction practice to the speculative and critical tradition in Dunne and Raby's work (Dunne & Raby 2013), Sterling's definition of 'the deliberate use of diegetic prototypes to suspend disbelief about change' (Sterling 2013), or Bleecker's influential book that especially elaborated the key concept 'diegetic prototype' (Bleecker 2009). With the two elements of narratives and speculation of the yet-to-exist, it is one of the valuable design attempts in investigating social, ethical implications of future technology. Thus, the value of this method is more than in generating imagination and novelty of technology. Instead, it is attentive to provoking discussion on sensitive and conflicting issues around the emerging technology. This is the 'speculative turn' in contemporary design practices called by Hales (Hales 2013): 'it creates a discursive space within which new forms of cultural artefact (futures) might emerge'. We also see it as 'a discursive turn' (Lindley & Coulton 2015) that design fiction is used as a research tool to ask better questions than providing technological solutions.

In design, the process of creating and using Design fiction is often participatory (Lyckvi, Roto, Buie, & Wu 2018). Design fiction is carefully presented to engage participants to provoke discussion, opinions or even further speculation and imagination. Thus, accordingly, design fiction takes a wide range of material and experiential formats. Markussen & Knutz (2013) describe a variety of multi-media as 'packaging' for design fiction stories. For instance, they are well-crafted exhibit objects and multiple media in museums (Auger 2013), functional artifacts brought to real everyday use settings of field (Pierce 2019; Søndergaard & Hansen 2018), collages and cardboard mock-ups used in co-design workshops (Hanna & Ashby 2016; Huusko, Wu, & Roto 2018), and experiential events and performance (Candy & Dunagan 2017; Elsden et al. 2017). Designers also borrow widely used daily items like advertisement poster (Bleecker 2014; Blythe, Steane, Roe, & Oliver 2015), commercial product catalogue (Brown et al. 2016), and products sold in 0.99 dollar grocery (Montgomery & Woebken 2016).

All these design cases have well illustrated that design, with its constructive tradition, lies the competence in bringing invisible, intangible futures to live, material or experiential forms for people to see, experience, comment on and interact with (Candy & Kornet 2019). And at the same time, such engagement allows researchers to study the emergent phenomena ethnographically for further enquiries (Lindley, Sharma, & Potts 2015; Smith et al. 2016).

**2.2 Design Fiction in Elder Care**

In the context of designing for elder care, smart and IoT technologies of assisting and monitoring are introduced. More people realise that design is not only about applying smart technology to ensure safety and health, but about translating technology into the values that matter to older adult individuals and their social environment (Leong & Robertson 2016). How would the older adult users' intimate experiences, psychological emotions, cultural values, and social connections be mediated by future technologies? To address this question, design fiction has been used relating to the domain topics like positive ageing in care centres (Blythe et al. 2015), assisting death (Tsekleves et al. 2017), volunteer services (Blight & Wright 2006), and dementia (Darby & Tsekleves 2018), (Noortman et al. 2019).

Carrying the discursive stance, design fiction is often used collaboratively with older adult participants in workshops. The older adult are either invited to co-compose and -develop fictions to express their desires or fears (Ambe, Brereton, Soro, Buys, & Roe 2019), or to comment on written fictions that were carefully crafted with provocative design concepts and plot by researchers or experts (Ahmadpour, Pedell, Mayasari, & Beh 2019; Tsekleves et al. 2017). A recent one takes a more performative and interventionist approach (Noortman, Schulte, Marshall, Bakker, & Cox 2019). Researchers made a probing prototype of remote care device and invited participants to play the role of caregiver for a fictional character of Annie living with dementia.

From those discussions, the recurring themes are mainly around the conflicting relations between the control imposed by technology and the craving for individual autonomy and independence (Soro, Ambe & Brereton 2017). Ambe and her colleagues (2019) reported that their older adult participants expressed a strong desire for breaking from the 'invisible power of watchman' for adventure regardless of the dangerous results. Schulte (2016) portrayed 'a stubborn father' who refuses to wear monitoring technology. In the short video 'Uninvited guests', a 70-year-old man gets frustrated by the monitoring ecosystem and plays tricks to treat the system (Superflux 2013). They are all strong-minded individuals who succeed in gaining (back) power and control in the end and value things like home comforts and a sense of autonomy.

These types of characters imply cautious and reflective attitudes towards 'temptations of technology' as IJsselsteijn et al. (2020) call it. To give several examples, the hypothesis that made most of the design projects fundable is the perception of technology as the solution to most problems of elder care. However, we shall ask whether the technology is actually relevant to the problem space at all. Regarding the solution provision, sensors are installed everywhere unquestionably to track, monitor, and respond. Who is then actually benefiting from these monitoring systems, the one doing the monitoring or the older adult?

# 3 Crafting futures of elder care in Singapore

The design fictions in this study are a spin-off from a three-year long government-funded project that explored future nursing home typologies in Singapore. The research project included a design exploration component where the team, comprising architecture and design researchers, ideated, and proposed future concepts of care. The future proposals centred around two themes that were of particular interest to the funding agencies - care automation using robots and integrating eldercare services into public housing - for their potential to alleviate staffing challenges in the eldercare sector and to strengthen the integration of nursing homes with the community. While the potential of robots to provide utility seems promising, having a robot at home is far from the everyday reality of older adult Singaporeans, who grew up in an era of low technology proliferation. Robots' social acceptability in caregiving has also yet to be thoroughly explored, especially in the home setting where these technologies interface with family values and dynamics informed by culture. Moreover, locating nursing homes within public housing estates have created tensions with the residential community in the past, which could prove a barrier to future integration.

In the two design fictions presented here, we turned these nuanced issues to the future world five to ten years ahead, where the proposals in the future concepts have become 'true' or widely practiced.

The design fictions were hosted on a free website. The sites featured mock-ups of real-world websites - Singapore's major newspaper ('The Straits Times') and the online shopping site ('Lazada') to create a make-believe setting. The links for the mock-ups were distributed through social media, where participants were informed of the purpose of our research and our intent to gather their thoughts on the fictional material presented to them via the links. Upon landing on the site, participants had to acknowledge an automatic pop-up disclaiming the fictional nature of the materials before they can continue to view the site.

### 3.1 Design Fiction 1: Automating Care at Home "Give your parents the gift of care this Chinese New Year!"

This fiction explores possible tensions between two phenomena observed. One, is the government's push towards automation and smart technologies to cope with the projected rise in demand for eldercare services. Robots for caregiving has been one of the key development areas driven by the Singapore government (Singapore is turning to artificial intelligence for elder care 2017), (Assistive Technology and Robotics in Healthcare n.d.), followed by local research institutes (AboutCHART n.d.). Second, is the voice from family members visiting their parents in nursing homes, interviewed in the project. They carry feelings of guilt from not committing their responsibilities as children. This observation led us to question: when adult children feel guilty about placing their parents in a nursing home, how would they feel about entrusting their parent's care to automated technology such as a robot? How would it affect family dynamics in a culture where family ties and filial piety are stressed? To delve into these questions, we imagined and crafted a near-future world where robot caregivers are commonplace options to provide care for seniors at home, taken as just another household electrical appliance.

This fiction is presented through two fictional news articles conveying the government's push for citizens to adopt automated technology to care for their aging parents through subsidies (see Figure 1). The news article site shows an advertisement banner leading to a fictional product, an elder care robot in this case, listed on an online shopping site (Figure 2). This fiction was positioned as the practice of gift-giving to immerse the audience with the role of an adult child and provoke them to consider how they would feel buying a robot to care for their parents. Participants see a pop-up advertisement positioning the fictional robot as a "gift of care" to parents during Chinese New Year, a season when adult children give red packets to their parents as a symbol of their love and filial piety

The fictional robot was created based on technology analysis conducted on various robot cases over the course of the future nursing home project. Among those referenced are robots assisting care giver's physical work such as Robear (Dredge 2015) and Care-o-Bot (Care-O-bot ® 3 n.d.), and social robots providing emotional care to older adults such as PARO (PARO Robots 2014) and Dinsow (Kishimoto 2017). The visuals were created by modifying and combining existing images to create a believable fictitious product.

After exploring the online shopping page, participants were guided to a form to key in their answers to their fictional decisions. Participants were first asked, "would you like to buy Rumii for your parents in their old age?", then they are guided to share their thoughts on two questions - "Why do you want / not want to buy Rumii for your parents in their old age?" and "How do you think it may affect your relationship with your parents?".

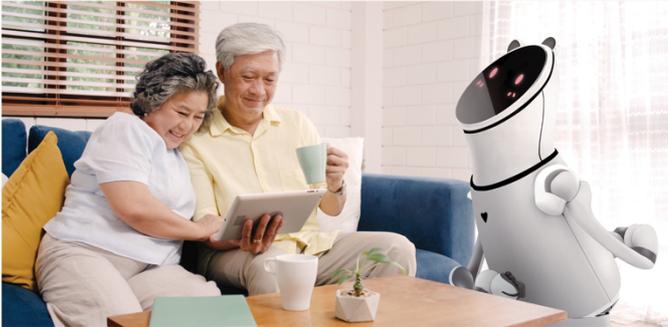

# THE STRAITS TIMES

02 | MONDAY, APRIL 17, 2025

## $500 Senior's Tech Support Grant for Singaporeans to buy Rumii Care Bot for Ageing Parents

SINGAPORE - Singaporeans with at least one parent aged 60 and above are eligible for a one off $500 grant when purchasing the Rumii care bot to care for their ageing parents. The care bot retails at SGD 9,000. Singaporeans simply need to show their proof of purchase and proof of registration of the care bot to a family member above the age of 60 to claim the grant. This can be done via the One Service Mobile App or at the nearest community centre.

These initiative is rolled out to encourage adoption of the Rumii care bot as part of the nation's push towards ageing-in-place for senior citizens. This initiative is expected to help reduce demand for care home placements, which had steadily increased over the years following a easing of the placement eligibility criteria, and is currently fast outstripping supply, as more are unable to keep up with the stresses of caring for both their elderly parents and their children.

The government hopes to achieve an adoption rate of 70% or more among seniors and their families. The eligibility age for the grant has been set at 60 years old to encourage early adoption among the young old, as preliminary research findings indicates that Rumii could improve health outcomes when adopted while seniors are still healthy.

### Rumii Care Bot
★★★★½ 212 Ratings

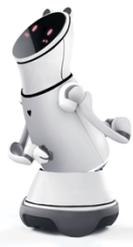

**CNY SALE!**
Now only
**$7,888.88**
U.P $9,000.00

Want to show your parents how much you love and appreciate them? Get Rumii.

*Scan to view the Design Fiction site!*

SCAN ME

## Can an AI Care Bot be the Panacea to the Woes of the "Sandwich Generation"?

SINGAPORE - There has been much abuzz over the past few months about a first-of-its-kind, made-in-Singapore care bot called Rumii. It is anticipated to be a game changer in the eldercare sector in helping to relieve family caregiver strain. A pilot trial was completed with 250 early adopters representing a range of care needs, including seniors living alone in need of care, who otherwise would have been admitted into care homes. While the pilot programme was met with generally positive feedback, there was also backlash, with critics arguing that the advent of Rumii will enable society to "outsource" their elderlies to robots and fuel elderly neglect, especially among those with strained relationships.

*Figure 1. News articles from design fiction 1: "Give your parents the gift of care this Chinese New Year!"*

*Figure 2. Online shopping page with fictional customer reviews from design fiction 1: "Give your parents the gift of care this Chinese New Year!"*

## 3.2 Design Fiction 2: Community-based care "Join the Petition to bring integrated care services to our residence!"

In the past, the announcement of plans to build a nursing home had elicited resistance from residents of the neighbourhood. This community's NIMBY, i.e., "not-in-my-backyard", attitude towards nursing homes received significant media attention back in 2012, with one resident who opposed the idea even commenting that "the old folk will be groaning right into my home" (Seow 2017). In our project, we proposed a future model of care where care services are integrated into the public housing estates called HDB (Housing Development Board), where more than 80% of Singaporeans currently live. In this scenario, community members are leveraged upon to provide social support. This future concept proposal was targeted to meet the government's current push towards ageing-in-place. Given the history of negative public perception towards eldercare facilities, however, we wanted to explore people's responses by inviting them into this future scenario. We imagined a potential future where such integrated care neighbourhoods, called "Care Corridors" are being piloted by the government, and have garnered enough popularity that neighbourhoods are petitioning to have Care Corridors built in their estate, a reverse of the NIMBY situation - "Please-in-my-front-yard" (PIMFY).

In the Care Corridors fiction, we explored the dilemma of balancing autonomy and safety, which we observed in local and overseas nursing homes, especially for residents with dementia. Currently, care staff grapple with balancing the two, sometimes using restrictive strategies to monitor residents and minimise safety risks, such as keeping residents seated in a visible area or using restraints. In a community setting, neighbours in the estate can be leveraged upon to provide a collective watchful eye and safety net but are likely to face a similar dilemma. How will the community respond?

Participants are presented with a fictional petition site, inviting the public to vote for the government to develop integrated eldercare services at their public housing estate. The petition presents visual and written details of the amenities and programs available, with a particular focus on the community's contribution to care such as intergenerational co-housing and crowdsourced 'safety monitoring' for older adults with dementia as key selling points. At the end of the webpage, participants were guided to a form where they are invited to consider how they felt about such an arrangement through a series of questions. Participants were asked to make a fictional decision whether to support or oppose the petition, then guided to share their thoughts on two questions – "Why do you support/oppose this petition?" and "How do you think the Care Corridor will affect life in your neighbourhood?".

Both fictions present the fictional comments by the citizens who live in each future world as if they responded to 'Buy or not' within the online shopping page and 'Vote or not' in the petition website. The comments were crafted based on insights gleaned from field research during the nursing home project regarding the lived experiences of persons with dementia in care homes, the struggles faced by caregivers in caring for a loved one with dementia and the community's perception of eldercare facilities being built in their neighbourhood. Presentation of those comments was targeted to provide varied and contrasting perspectives of complex needs and issues.

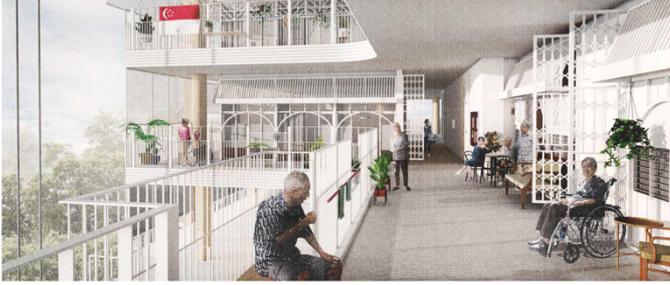

Figure 3. Design Fiction 2: Community-based care "Join the Petition to bring integrated care services to our residence!"

## 4  Putting the Fictions Out There

We published the two design fictions on online websites and allowed the links to circulate organically on social media channels such as Facebook, WhatsApp, and Telegram interest groups to invite the public to experience and interact with the fictions. We collected the participants' responses through online commenting, as if they were making decisions of whether to buy or support in the two future settings, as explained above. In the online commenting, they were asked to submit their age and gender, with an option of non-submission, by being informed that the data is collected only for the research purpose.

*Table 1 Respondent breakdown according to age range and gender for Design Fiction 1 (DF1) and 2 (DF2)*

|  | Design fiction 1 | | | | Design fiction 2 | | | | Total |
|---|---|---|---|---|---|---|---|---|---|
|  | Female | Male | No submit | DF 1 Total | Female | Male | No submit | DF 2 Total |  |
| 18-24 | 7 | 1 | 0 | 8 | 4 | 1 | 1 | 6 | 14 |
| 25-34 | 28 | 8 | 1 | 37 | 26 | 8 | 1 | 35 | 72 |
| 35-44 | 3 | 1 | 0 | 4 | 3 | 1 | 0 | 4 | 8 |
| 45-54 | 2 | 0 | 0 | 2 | 1 | 0 | 0 | 1 | 3 |
| 55-64 | 4 | 1 | 0 | 5 | 6 | 0 | 0 | 6 | 11 |
| 65 and up | 0 | 0 | 0 | 0 | 1 | 0 | 0 | 1 | 1 |
| Total | 44 | 11 | 1 | 56 | 41 | 10 | 2 | 53 | 109 |

In total, we collected 56 responses on the first fiction and 53 responses on the second. While the participants' ages vary in a range from early 20s to over 65 years old, many of them fall within the 25-34 age group (66% of the total 109 responses) followed by the 18-24 age group (13%) and the 55-64 age group (10%). 85 out of the 109 responses (78%) were from females. The bias towards the younger age group can be due to the limitations of our chosen medium of online websites and fictional stories. Participants in the younger age group, being more accustomed to interacting with content on social media, are more likely to be engaged in going through the entire fiction and leaving their responses. In addition, this age group will likely be the ones needing to buy care robots for their older adult parents in the timeline pictured in the design fictions. Significantly more female participants commented in response to the fictions (85 of 109 participants). The authors speculate this may be due to women experiencing higher levels of relatability to the topic and so being more motivated to comment, reflecting the cultural expectation for women to bear the burden of caregiving as 60% of informal caregivers in Singapore are female (Chan 2010).

We conducted a workshop to interpret the written responses based on a grounded theory approach (Glaser & Strauss 1967) and identified several themes depicting the tensions. Based on the themes, we tweaked the design fiction visuals presented in this paper to represent our interpretation of the participants' contributions. This was intended to make the entire design fiction process reciprocal and dialogical between the researchers and the public, which can imply further fiction creation and issue investigation.

## 4.1 Reactions to Design Fiction 1

### 4.1.1 Negotiation of Labour Division between Robot and Adult Children in Caregiving

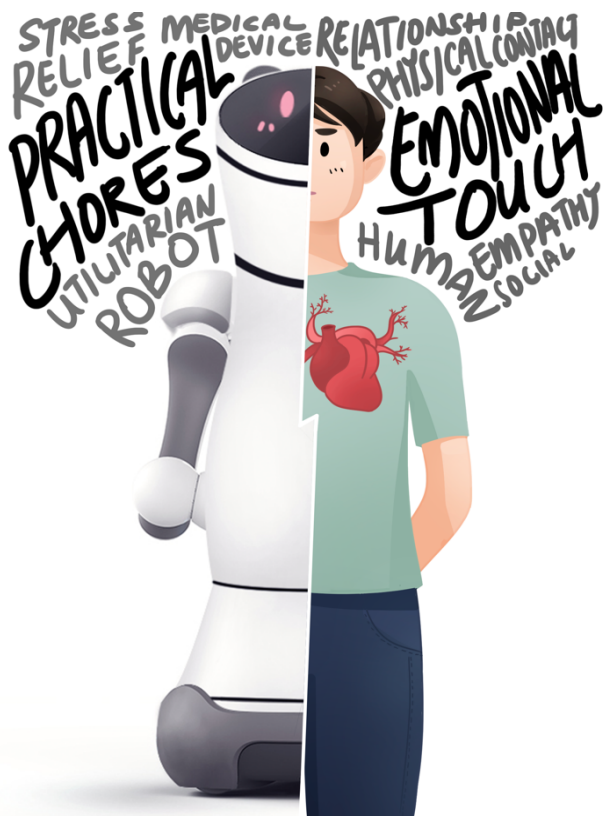

*Figure 4. The figure illustrates the participants' perceived division of labour between robot and human caregivers, with the robot serving a utilitarian and practical function, while they themselves fulfil the emotional and relational needs of their parents which the robot cannot replace. (Illustration by the first author)*

The participants' comments reveal the negotiation of the care work between the care robot and the adult children, where the care robot's function is delegated for practical chores while humans are responsible for providing the emotional touch. Participants preferred the care robot to fulfil practical care needs. They conceived that the robot would relieve the stress of performing practical caregiving tasks and help them achieve the ideal intimate and caring relationship with their parents, which was to spend quality time together socially and engage emotionally.

> "Our conversations may be less naggy/trivial to more interesting conversations if rumii handles their basic needs so that I don't have to attend to those." (Female, 18-24)

> "It may take off the stress and leave me time to spend relaxing and having fun with my parents." (Female, 25-34)

The comments also reveal an underlying competitive dynamic between the adult children and the care robot. The roles played by the two should complement, not overlap, to maintain a good balance in the care relationship with their parents.

> "...it'd be better if Rumii was less personal - to clearly demarcate the difference between a utilitarian robot vs one that comes off as too empathetic and appears to be a replacement for human touch." (Female 25-34)

"I think it depends on how my parents take it and how well I can create a good dynamic between the robot's role and I (that it doesn't feel like the robot takes over me but the robot is used to make life better for them)" (Female, 25-34)

"I only need Rumii to manage daily schedule and routine, and light engagement. While for deeper engagement, my family and I can fulfil that." (Male, 35-44)

### 4.1.2 Social Pressure for Both Children and Parents

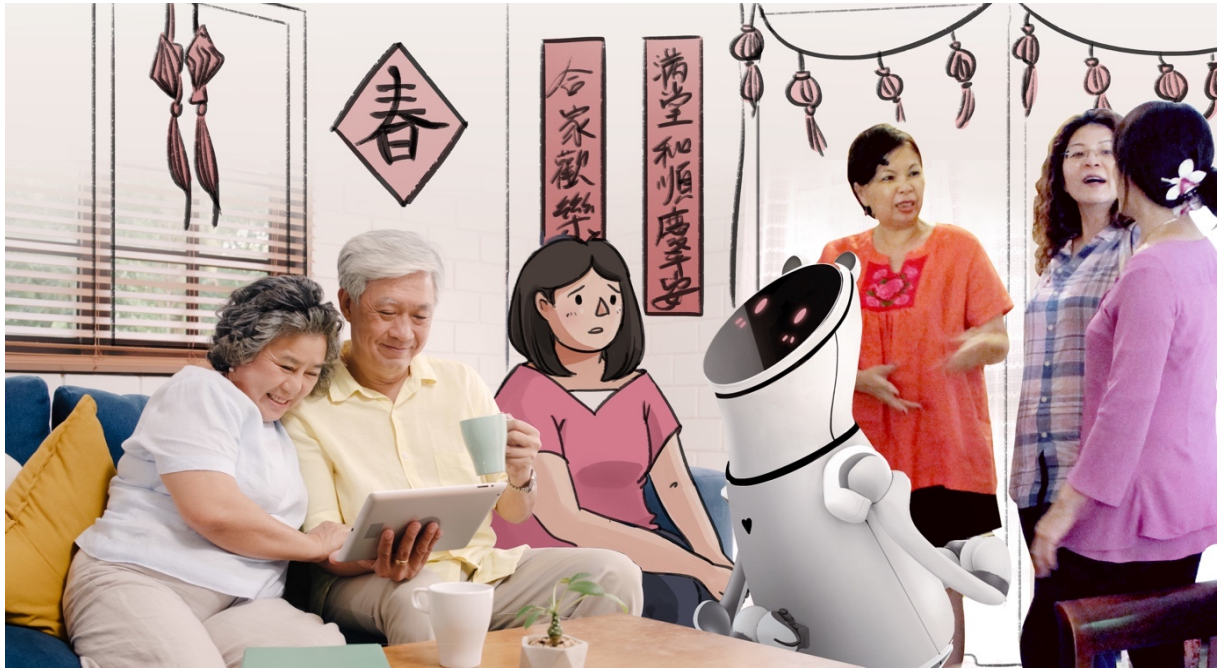

*Figure 5. The figure illustrates an imagined scenario where relatives visit the participants' home during Chinese New Year, where the care robot can be seen tending to their parents' needs. The participant's persona sits in the centre worried, as though wondering "what will our relatives think of us?". (Illustration by the first author)*

We found that participants' comments valuing providing care for their parents not only manifests their personal will but also their concern for their social image to their relatives. Interestingly, their concern on their social image appears twofold. First, they worried about their social image that might be portrayed as irresponsible children to their elder parents. Secondly, they also worried that their parents' social image would get negatively affected as well as they brought up irresponsible children.

"I feel like it will give a negative impression to the other relatives if I buy Rumii for my parents." (Female, 25-34)

"I also think they may not want relatives to know that they rely on a bot for companionship. Dunno, for me somehow if it is a robot to check their vitals, maybe it gives my parents a high-tech image, but if it is for social companionship then it seems embarrassing" (Female, 25-34)

### 4.1.3 Tension between Pragmatism and Traditional Chinese Values

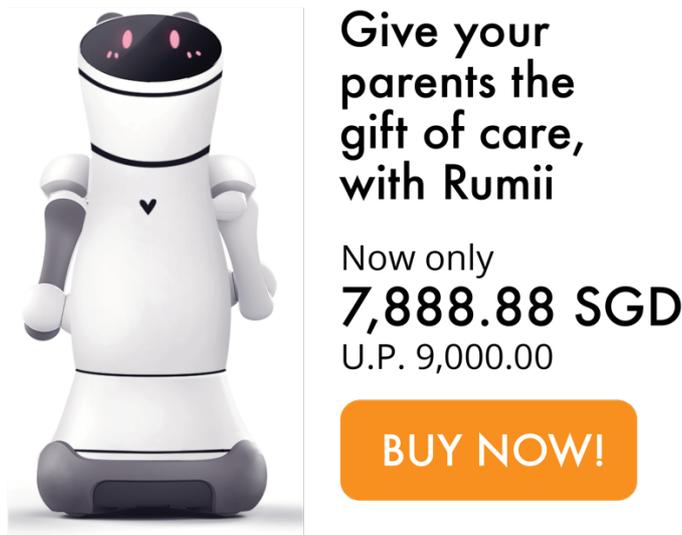

Figure 6. The "Value Calculator" illustrates the weighing of practical benefits and emotional trade-offs that participants grapple with when considering purchasing the robot caregiver. (Illustration by the first author)

Many responses projected the worry that buying a care robot for parents might give their parents the impression that children want to replace their care responsibility with the robot. Their comments reflected their concerns of disappointing their parents' expectations, the feeling of guilt, negative social image, and their over-reliance on technology resulting in less interaction with their parents. Some comments were clearly against the idea of robot care as their parents "would be insulted" (25-34, Female). Some other comments did think of the robot's practical benefits such as health monitoring but still appear to place more priority on pursuing the ideal image as responsible children.

> "There'll definitely be some form of disappointment. This is especially because of the expectations Asian parents have on their children, that they have to look after them when they are old. However if the use of the robot has been normalised and they see that many of their friends use it, it might be more acceptable." (18-24, Female)

"It's the children's responsibility to care for their own parents and not outsource this to a bot. Similarly, my parents didn't get a bot to care for me when I was a baby." (Male, 35-44)

"Now I see that it'll be really easy for them to feel like I've bought this so that I can abandon them. I can imagine my parents pretending to like the bot, just so that I'll feel okay with leaving them with it. Knowing my parents, they don't like to feel like they are a burden on us." (25-34. Female)

On the other hand, there were also comments that reflect the participants' practical perspectives in evaluating the robot's worth, by calculating its price against its functions. Some mentioned that the robot was expensive, and they might consider if it was cheaper.

"No, it's still too expensive after a mere $500 subsidy. It should be subsidized by at least 80%." (45-54, Female)

"Anything that can improve the lives of my parents is worth a consideration. But my main issue is the price. That's really steep." (Male, 25-34)

"Price is a little steep, not sure if people will be convinced that its functionalities are worth the $." (25-34, Female)

## 4.2 Reactions to Design Fiction 2

### 4.2.1 Future Projected Empathy and Reciprocity

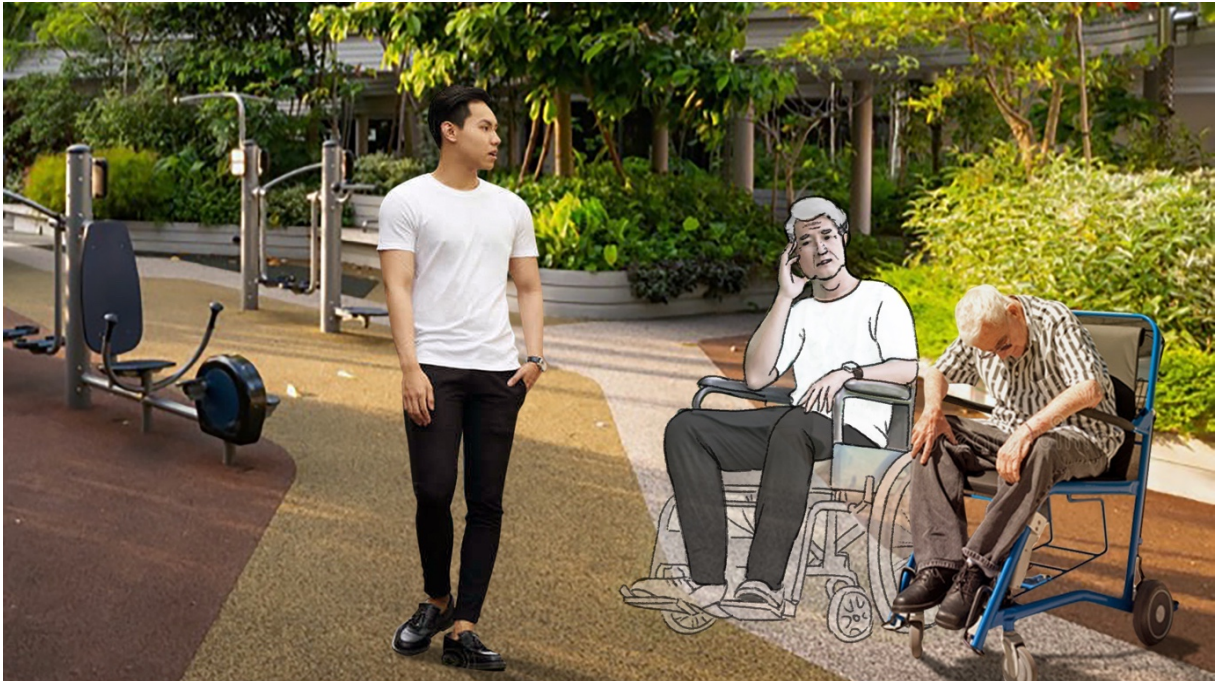

*Figure 7. The author imagines the young respondents putting themselves in the shoes of the older adult residents in the fictional care corridor. (Illustration by the first author)*

The responses showed overall positive support to the fictional petition. The majority of young respondents supported the petition and regarded themselves as possible caregivers although almost none explicitly stated the benefits they would get to their current life or seriously considered the intensive time or energy they might need to invest. Mainly, they developed the type of future projected empathy that if they treat the older adult well now, when they get old, they will get treated well equally. One particularly mentioned that he did not want to be isolated when he gets old, therefore, it is not nice to isolate older adults with dementia.

> "We will be old one day, and we would not want to be isolated in our small and dark room." (Male, 35-44)

> "It helps to foster a stronger and healthier community spirit by helping one another, especially since we would all be old someday." (Female, 18-24)

> "We have to take some responsibility in caring for the older generation as we will be them in the future and will face the same problem." (undisclosed gender, 18-24)

### 4.2.2 Prioritising the Role of Good Citizens

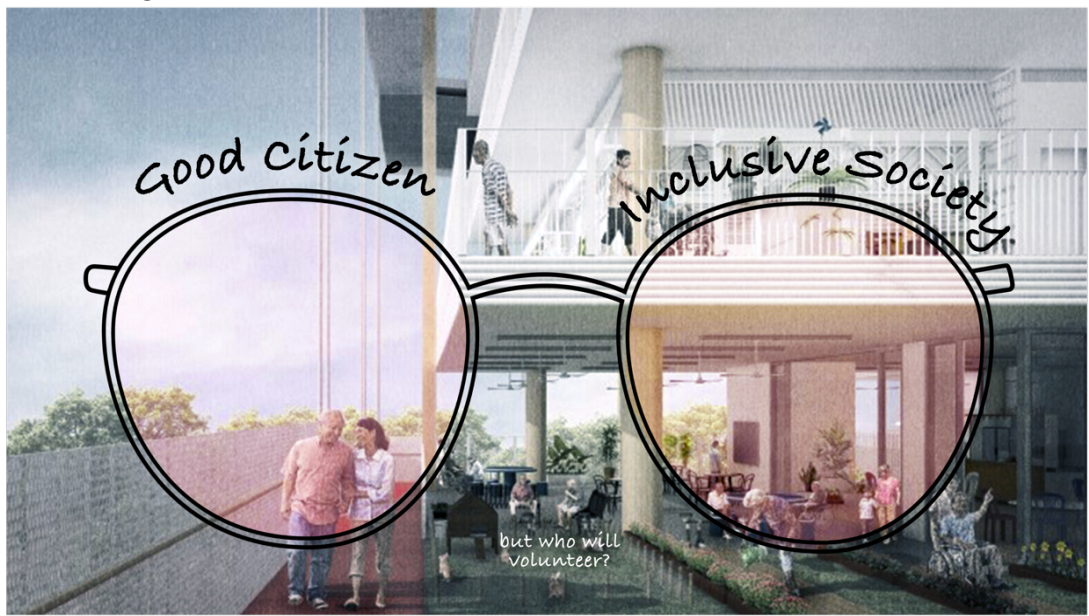

*Figure 8. The participants respond from the perspective of idealistic good citizens, thus viewing the situation through "rose-tinted glasses". (Illustration by the first author)*

When commenting on the petition, participants seemed to take a role as a good citizen championing the ideal of a good and inclusive society, where community members take care of one another. In their comments, participants appeared taking a distance from what might likely happen, instead of thinking of themselves as a community member providing care, as reflected in one comment, "I'm kinda just curious to see what it would be like haha, and how it would affect the community spirit/vibe in my neighbourhood" (Female, 24-35). Potential tensions that may arise from being a community member contributing to the care of seniors in the neighbourhood were barely reflected in the responses.

> "It's a more humane way to support the growing older adult population, especially those who are able enough to live outside of the nursing home system but are still looking for some form of care support." (Female, 25-34)

> "It's a brilliant idea! I think it's a wonderful way to foster an inter-generational and inclusive community." (Female 25-34)

> "Sounds good to have more people watching over older adult…" (Female, 25-34)

# 5 Discussion

## 5.1 Unpacking Tensions in Cultural Perceptions in the Asian Context

We unpacked key conflicting values of cultural perceptions from the creation process of the design fiction and responses from public participants. The first relates to the first fiction where care robots introduce tension into existing social dynamics in Singapore built upon the Confucian value of filial piety. Participants tried to negotiate the new relationship between themselves as the supposed caregivers, the robot as the new caregiver and their elder parents. They chose to uphold the ideal of being a filial child over the care burden the robot might reduce. Also, the more the robot takes on a social and empathetic role, the more it is viewed as a threat to the ideal filial child-parent relationship. This ideal is also constructed from social pressure, as reflected in the comments worrying about their social image getting damaged by using the robot.

The second value conflict lies in the coexistence of differing values in the current Singaporean society, which are pragmatism like cost calculation and the traditional Confucian values such as righteousness, loyalty, propriety, and filial piety (Kuah 1990). Pragmatism has been a national ideology instilled in the everyday life of Singaporeans, but also interacted with the Chinese Confucian values (Tupas 2015). As a result, in contemporary Chinese Singapore families, the values have found new meanings into pragmatic actions towards older adult such as efforts maintaining family harmony and social image, giving physical and financial care, spending time together and others (Mehta & Ko 2004), (Mehta & Leng 2006).

Thirdly, more evident in the second fiction, we see the dynamic relationship between personal concerns and benefits and the expectation of an ideal society. When deciding to support the fictional petition, young participants put the hard work of daily care for older adult residents with dementia aside. Instead, they prioritised the vision of a good society and took the role of good citizens supporting the values of respecting seniors, inclusivity, and multi-generational integration.

## 5.2 Social Relationship as the Protagonist in Elder Care Fictions

In investigating cultural perceptions of elder care in the Singaporean society, our work adds an Asian perspective to the overall future speculation of elder care. Most of design fictions on elder care and related discussions have been conducted in the western context. The clear pattern shared in those works is that the protagonist is the individual user who struggles with the tension with technological devices in use, implied by the typical User-Centred Design sense (e.g., see (Ambe et al. 2019)). However, in our work rooted in the Asian context with the dominant values of family and community, the protagonist in our fiction is the social relationship, either between the adult child and older adult parents or between younger and older adult residents. The being of the older adult is rather mutually constituted by the relationship with their adult children. And the losing value or worry caused by new technology is not only about the derived autonomy of individual user, but also the potentially damaged social image of the whole family caused by dismissing filial piety. Such plots of our design fictions reveal the social features of the Asian society drawing on our ethnographic observation from the future nursing homes project, revealing family members' feeling of guilt in not fulfilling their responsibilities as children by placing their older adult parents in nursing homes, and the general negative social stigma around the issue (Seow 2017).

In terms of engaging people with the fiction, we constructed the fictional roles of the older adult's adult children and neighbours instead of the older adult themselves. Although some Design fiction work considered relational and social aspects in creating fiction (Blythe et al., 2015; Schulte et al., 2016; Superflux, 2015), the social relationship is just one of the many qualities of older adult wellbeing instead of the focus. Especially when crafting the friction and conflicts, the focus is on the negotiation between the older user and technological machines. Similar technique can be found in Noortman and her colleagues' work who gave the research probe to the caregiver of the older adult patient (Noortman et al. 2019). However, we explored the social layer of the caregiving act which is not bounded by payment or unemployment but by kinship or neighbourhood. Another differentiation from related work is our study intentionally downplayed the friction with the machines. For instance, the one customer review from an older adult in the first Design fiction mourned the loss of connection and quality time with his/her son while at the same time praising the excellent value and performance of the care bot. Here, the conflict underlying social relationship was highlighted as elder was very dissatisfied regardless of the satisfaction with the technology. However, it also indicates the limitation of our work that issues and matters of concern could have been the interplaying result of the two social and technological aspects. Overall, we would argue that this study might provide the fundamentally distinguished setting and structure for fiction making on elder care in the global scene.

Relational quality at the aspects of family relationship and community structure has been always one key element in designing for elder care with dementia (Morrissey et al. 2017; Sabat & Lee 2012; Tsekleves et al. 2017; Vreugdenhil 2014). And there are substantive design practices that use social relationship as the locus of design brief or objectives (Houben et al. 2020; Muñoz et al. 2019; Wintermans et al. 2017). We would like to suggest further work in Design fiction to put more effort in portraying tension related to social relationships or adopting the strategy of making a piece of relation as the protagonist instead of any individual. This increasing recognition on the social might benefit the investigation of designing for future elder care as care is a fundamentally relational and responsive act which is beyond the discourse of autonomy and human-machine interaction.

### 5.3 Different Entry Points to Interacting with the Fictional World

We find participant responses from the first fiction appear more diverse with uncertainty and decision-making dilemmas while responses from the second appear more consistently supportive. The difference might be related to the different role-takings of participants according to the invitation formats. The first fiction invited the roles of consumer and the child of older adult parents, whereas the second invited the role of citizens supporting or opposing the petition. Participants used different references while taking different roles in both cases. As a 'child', they used mundane materials from real-life experiences relating to personal parent-child relationships. As a 'citizen', they used the moral value contributing to the community's common good than actual concerns or dilemmas relating to day-to-day care work. We speculate that the discussion or responses might be different in the second fiction of 'Care Corridors' if the invite asks for the role of caregivers, like 'Would you sign up for the volunteering work as a caregiver, and in which way'? This observation suggests a careful consideration when designing for the interactions between participants and design fictions.

The props and act related to consumption is a common engagement technique in Design fiction (e.g., Brown et al. 2016; Montgomery & Woebken 2016). Our study again has illustrated that the commercial purchase page is a valuable tool to get people at the present quickly immersed with the fictiveness, easily understand the content, and express their opinions based on their mundane lived experience. Moreover, our study took a step further that we used the opinions collected from the participants to continue the craft process of Design fiction. We developed more fictional scenes, for instance, the scene of relative visiting (in Figure 4), as a new entry point added to the world building and discursive space.

### 5.4 Challenges of Crafting Design Fiction and Suggestions for Future Work

We faced challenges balancing between proposing ideas and provoking with ideas. Our initial storytelling approach proved overly prescriptive and emotionally laden, guiding audiences what they should feel about the issue. We then moved to a 'World building' approach that focused on providing space for participants to construct their own views (Coulton et al. 2017). With this approach, we created everyday artefacts to give verisimilitude to the future concepts and to provide participants a role to embody as they formed opinions on the concepts presented. After collecting and interpreting participants' comments, we made visual tweaks to the design fiction materials, to project cultural perceptions and value conflicts arising from their responses. The tweaked fiction materials are our responses to their opinions, which aim to build continuous dialogue. As future work, the participants could be invited to a workshop where the tweaked fiction materials are shared and give rise to further discussions. The continuous discussion can also invite project stakeholders especially government agencies.

## 6 ACKNOWLEDGMENTS

We thank all participants who took the time to share their thoughts on the design fictions.